\documentclass[manuscript]{emulateapj}
\usepackage{graphicx}
\usepackage{color}

\shorttitle{LAMOST 1: A Disrupted Satellite in the Constellation Draco}
\shortauthors{Vickers et al.}

\begin{document}

\title{LAMOST 1: A Disrupted Satellite in the Constellation Draco}

\author{John J. Vickers\altaffilmark{1,$\dagger$}, Martin C. Smith\altaffilmark{1} \\ Yonghui Hou\altaffilmark{2,3}, Yuefei Wang\altaffilmark{2,3} and Yong Zhang\altaffilmark{2,3}}

\altaffiltext{$\dagger$}{johnjvickers@shao.ac.cn}

\altaffiltext{1}{Key Laboratory for Research in Galaxies and Cosmology, Shanghai Astronomical Observatory, Chinese Academy of Sciences, 80 Nandan Road, Shanghai 200030, China}

\altaffiltext{2}{LAMOST Builder}

\altaffiltext{3}{Nanjing Institute of Astronomical Optics \& Technology, National Astronomical Observatories, Chinese Academy of Sciences, Nanjing 210042, China}

\begin{abstract}

Using LAMOST spectroscopic data, we find a strong signal of a comoving group of stars in the constellation of Draco. The group, observed near the apocenter of its orbit, is 2.6 kpc from the Sun with a metallicity of -0.64 dex. The system is observed as a streaming population of unknown provenance with mass of about 2.1$\pm0.4\cdot10^{4}$ M$_{\odot}$ and brightness of about M$_{V}\sim$ -3.6. Its high metallicity, diffuse physical structure, and eccentric orbit may indicate that the progenitor satellite was a globular cluster rather than a dwarf galaxy or an open cluster.

\end{abstract}

\section{Introduction}\label{sec:introduction}

With the rise of massive spectroscopic surveys such as The Sloan Digital Sky Survey \citep{yor2000}, the RAdial Velocity Experiment \citep{ste2006}, the Large Sky Area Multi-Object Fibre Spectroscopic Telescope survey (LAMOST; \citealt{cui2012}) and others, new phase space information is becoming available on massive scales.

However, while spectroscopic surveys provide additional tools such as radial velocity profiles, chemical compositions, and surface gravities, new hindrances such as incomplete photometry and poorly defined parameter-space coverage need to be accounted for. Structure searches traditionally rely on basic star counting techniques such as: statistical photometric density contrasting (see for example \citealt{new2002}, \citealt{maj2003} and \citealt{bel2006}) and photometric matched filter analyses \citep{ode2001}. But since spectroscopic data are inherently incomplete, new methods must be used to uncover structure, usually by looking for velocity clumping, see for example \citet{sch2009} and \citet{sta2009}.

We trawl the LAMOST spectroscopic data for distinct clumpings of stars in both velocity and metallicity spaces which are incompatible with the background distributions. The technique uncovers numerous known objects and structures as well as one strong detection of an unknown object which is the subject of this letter.

We postulate that this object is a star cluster being observed near the apocenter of its eccentric orbit. It manifests as a signature distinct from the background in velocity and metallicity space. We conclude that the object is likely to be a nearby, intermediate age globular cluster which is being disrupted by the galactic potential.

In Section \ref{sec:data} we describe the data used in this search; in Section \ref{sec:searching} we describe the search algorithm, rejected candidates and the Draco clump as well as its significance; in Section \ref{sec:properties} we calculate the age and distance to the clump via isochrone fitting, the object's size, mass and luminosity, and its orbital parameters as well; we discuss and conclude in Section \ref{sec:discussion}.

\begin{figure}
\includegraphics[width=\linewidth]{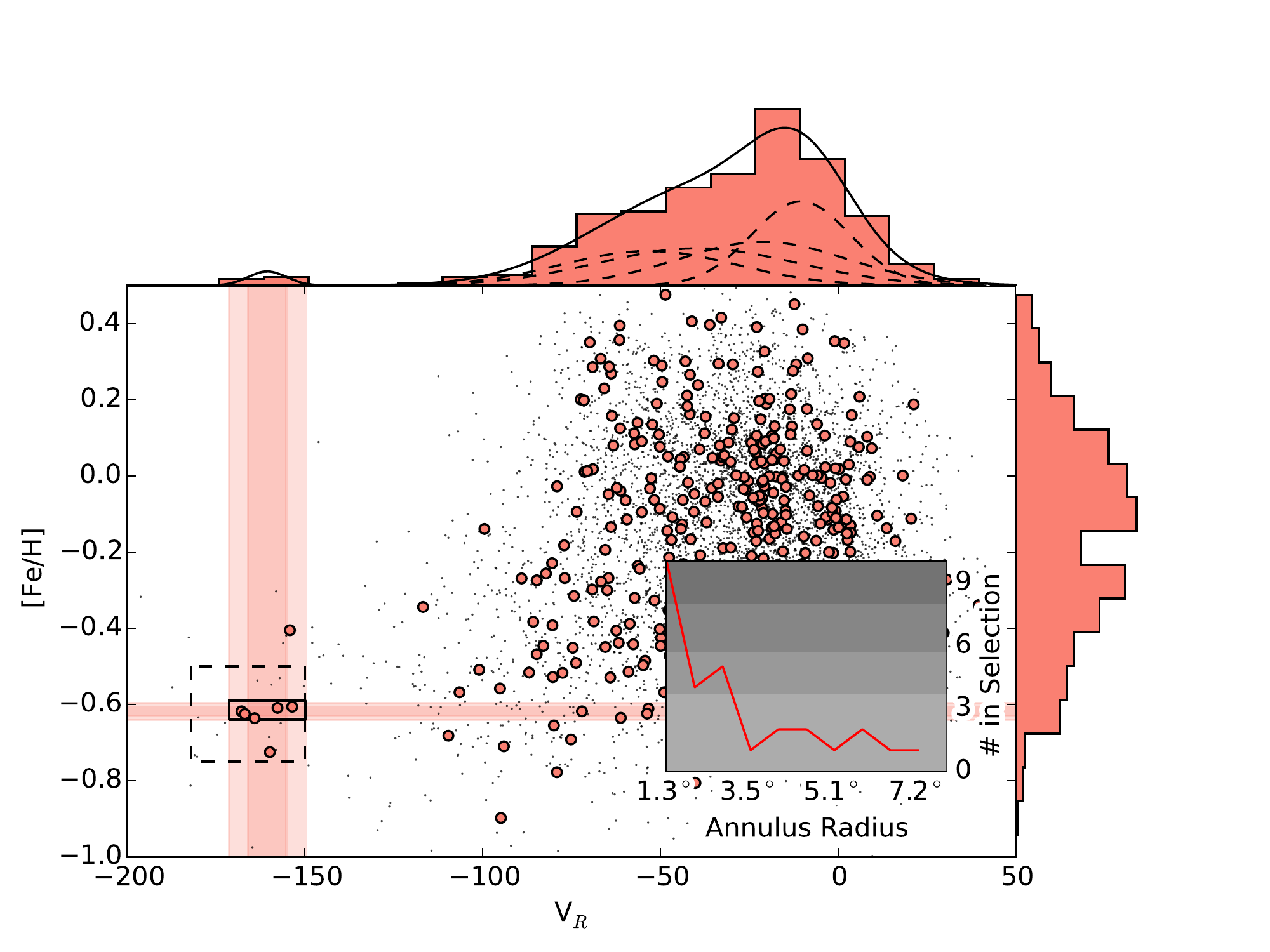}
\caption{Radial velocity and metallicity of LAMOST stars within 1.5$^{\circ}$ of ($\alpha$, $\delta$) = (274.5, 52.33) plotted as large circles and within 5$^{\circ}$ plotted as small dots. The top and right panel histograms indicate the velocity and metallicity distributions of the large circles (smaller sky area objects). \\
The solid selection box is defined algorithmically while the dashed box is selected by eye to increase the sample size of possible cluster members. \\
The inset plots the number of objects in equipopulated, 1000 item annuli which also fall into the dashed selection box as a function of average radius of those annuli. The shaded grey regions indicate the one, two, three and four $\sigma$ expectation levels for this value (calculated from a background annulus of 5$^{\circ}$-10$^{\circ}$).}
\label{fig:fehrv}
\end{figure}

\section{Data}\label{sec:data}

For our data, we use the $\sim$2.5 million stellar spectra available in the second data release of the LAMOST spectroscopic survey. The LAMOST telescope is 4 meter Schmidt reflector located at Xinglong Station, Hebei, China. The spectroscopic survey is performed by a novel system of 4000 individually positionable fibers which collect moderate resolution spectra (R$\sim$1800) in the optical range from 370 nm to 900 nm. For information on the telescope and survey, see \citet{luo2015}.

Since the LAMOST survey includes multiple observations for many objects, these data are duplicate cleaned with a pairing radius of 7" where the observations with the highest signal to noise ratio are preferred.

In Sections \ref{sec:sig} and \ref{sec:orbit} we will briefly investigate the kinematic properties of the stars using proper motions from the fourth edition of the United States Naval Observatory CCD Astrograph Catalog (UCAC4; \citealt{zac2013}).

\begin{figure*}
\includegraphics[width=\linewidth]{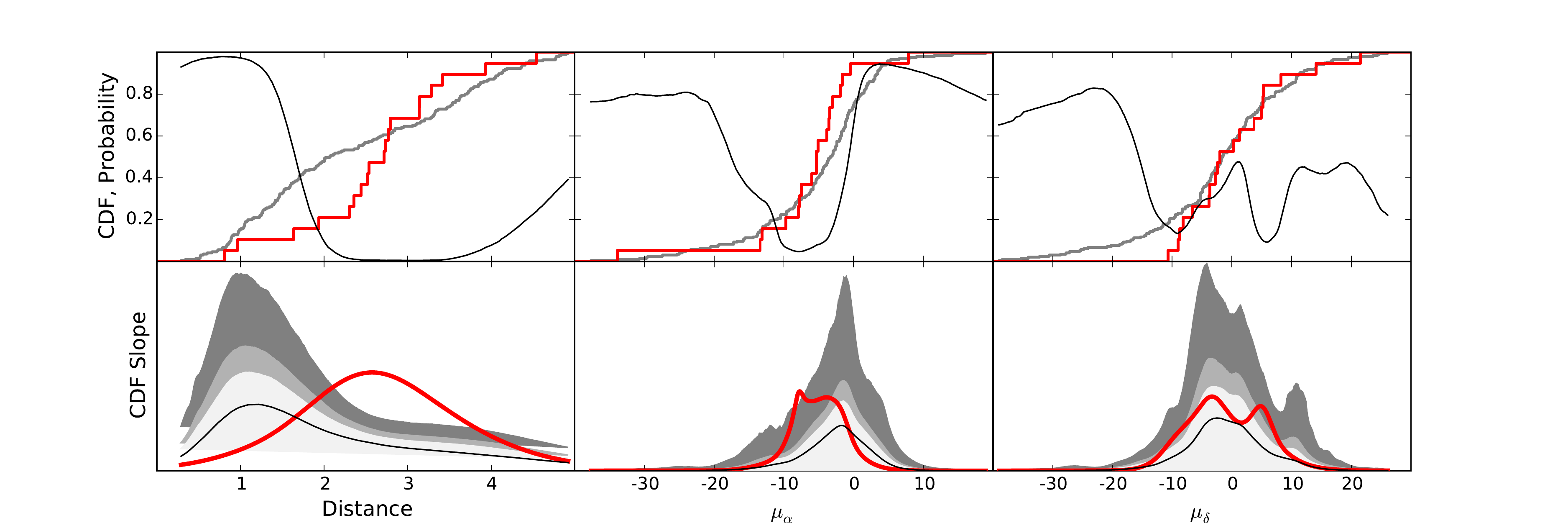}
\caption{The significance of distance (determined from LAMOST spectroscopy and photometry using a bayesian technique; from \citealt{wan2015}) and proper motion (from UCAC4) detections, using a technique similar to \citet{sch2009}. In the upper panels, the cumulative distribution functions are plotted for the target (in red) and the background annulus (thick grey). The lower panels show the slope of the cumulative distribution functions of the target (thick red) and the annulus (median in black, 75\%, 85\% and 95\% confidence intervals in shades of grey). The lower panel slope diagrams are smoothed according to the errors on the individual measurements. The confidence intervals are calculated by performing 10,000 random samples of the background population to mimic possible target observations. These Monte Carlo expansions also provide the total probability that the target population is drawn from the background population at each point along the CDF (thin black line in the upper panels).}
\label{fig:schlaufman}
\end{figure*}

\section{Searching}\label{sec:searching}

We search for substructure in the LAMOST dataset with one main constraint: the data are photometrically incomplete. With this in mind, we perform a gridwise search through the LAMOST spectroscopic dataset searching for metallicity-velocity clumping that is inconsistent with the field.

The search analyzes 1$^{\circ}$ radius pencil beams in 0.5$^{\circ}$ steps over the entirety of the LAMOST data set. The radial velocities of the stars in the pencil beam are then fit to a mixture of Gaussians with a varying number of component Gaussians (from one Gaussian to seven). If any component Gaussian has a velocity dispersion of less than 20 km s$^{-1}$ and has a minimum population of three objects (to avoid delta spikes), we check if that component is also two standard deviations distinct from the total field.

This search yields a large amount of stream-like candidates ($>$1000). We further reduce the candidate sample by investigating only those whose velocity outliers are also quite compact in [Fe/H] space, having metallicity dispersions of less than 0.1 dex. This reduces our sample to a more manageable 21 candidates.

We next compare the sky-coordinate positioning of these candidates to: 1) the globular cluster catalog of \citet{har1996}, 2010 revision\footnote{http://www.physics.mcmaster.ca/~harris/mwgc.dat}; 2) the nearby galaxy catalog of \citet{kar2013}; 3) the composite open cluster catalog of \citet{kha2012}, \citet{kha2013}, and \citet{sch2014}; and 4) the Sagittarius stream simulations of \citet{law2010}.

We find that, of the 21 candidates: four are coincident with the Sagittarius stream, four are consistent with M13, one is a known open cluster. Of the twelve remaining, three are single detections and six are exact duplicate detections in adjacent fields, we ignore these. The remaining three candidates are clustered in a three degree area, do not coincide with any known structure in the catalogs listed above, have consistent properties and are not duplicates. This overdensity is the subject of this letter.

In Figure \ref{fig:fehrv} we plot the velocity-metallicity plane for spectra in a 1.5$^{\circ}$ (radius) circle about the mean detection coordinates ($\alpha$, $\delta$) = (274.5, 52.33). These data are fitted to a mixture of Gaussians in velocity space, as shown. Then the metallicity of the structure is estimated via an iterative 2$\sigma$ rejection of the data falling into the outlying velocity profile. This algorithm selects the solid selection box shown.

A looser, dashed selection box is constructed by eye around the region: -182 km s$^{-1}$ $<$ V$_{radial}$ $<$ -150 km s$^{-1}$, -0.75 dex $<$ [Fe/H] $<$ -0.5 dex. This selection box is chosen to encompass all members in a larger sky area which appear as if they could plausibly be associated with the structure.

\subsection{Significance of Overdensity}\label{sec:sig}

\subsubsection{Significance in Metallicity Velocity Space}

To calculate the significance of the overdensity in metallicity and velocity space, we compare objects within 5$^{\circ}$ of the target field at ($l$, $b$) = (80.6, 26.1) to a background of objects in a local annulus of 5$^{\circ}$ to 10$^{\circ}$. Here we assume that the target selection function is equivalent in these two fields.

To remove the disk from our data sample, the fraction of which may vary between the two fields, we consider only stars which have radial velocities less than -100 km s$^{-1}$ and metallicities less than -0.5 dex\footnote{If we do not incorporate this cut, the total number of stars in our on-target and background annulus are 5998 and 11197, respectively. Repeating the significance calculation, we find that the probability of finding 24 stars in our on-target selection box is 1.3\%.}. Our on-target field has 204 stars in this velocity-metallicity range, of which 24 fall into the loose overdensity selection box described above. For the background annulus the numbers are 376 and 27, respectively, i.e. the probability of finding a star in the loose overdensity selection box is only 7\% and our on-target field is expected to have less than 15 stars. Therefore, using binomial statistics, the chance of obtaining at least 24 such stars from a sample of 204 is only 1.2\%. This constitutes our first probability.

This significance is shown visually in the inset of Figure \ref{fig:fehrv}. The frequency of stars falling into this selection box is plotted as a function of radius from the center of the detection and the significance compared to a background frequency is shown in shaded regions. The center of the group is almost 4$\sigma$ above the background and there are three \emph{independent} bins with detection over 1$\sigma$.

\subsubsection{Significance in Distance and Proper Motion Space}

To calculate the significance of the overdensity in distance and proper motion space, we refer to an analysis technique outlined in \citet{sch2009}. The essential idea is to compare the slopes of the cumulative distribution functions of two populations in some parameter space, a steeper slope in the cumulative distribution function relative to that of a smooth background indicates an overdensity at that point in the parameter space. This test is more attuned to local overdensities than the widely used Kologmorov-Smirnov test, which is commonly used to compare global distribution functions. By sampling the background distribution many times, confidence intervals may be inferred. This test is shown in Figure \ref{fig:schlaufman} and yields probabilities of 0.005, 0.05, 0.09\footnote{If we do not incorporate the disk cut described above, the results are unchanged.} for the distance, $\mu_{\alpha}$, and $\mu_{\delta}$ distributions (respectively) having been sampled from the background annulus described above.

\subsubsection{Total Significance}

Assuming that these probabilities are independent of each other, we combine them as follows:

\begin{equation}
P = P([Fe/H], v_{r}) \cdot P(\mu_{\alpha}) \cdot P(\mu_{\delta}) \cdot P(dist).
\end{equation}

\noindent finding a total probability of 3.0e-7 that it is drawn from the local background (5.1$\sigma$).

The 5$^{\circ}$ on-target field of our overdensity contains 204 stars, which is drawn from a total LAMOST population of 22,798 such stars (after the disk cut) across the entire survey footprint. Therefore our full sample contains approximately 112 similar-sized samples. The binomial probability of finding one or more detections at this magnitude in 112 random samples is approximately zero.

\begin{table}
\begin{center}
\caption{Properties of the Clump}
\begin{tabular}{|c|c|}
\tableline
R. A. & 274.5$^{\circ}$ \\
Dec. & 52.3$^{\circ}$ \\
$[Fe/H]$ & -0.64 dex \\
$\mu_{\alpha}$ & -7.8 mas yr$^{-1}$ \\
$\mu_{\delta}$ & 6.0 mas yr$^{-1}$ \\
Distance & 2.6 kpc \\
Age & 11.0 Gyr \\
\tableline
\end{tabular}
\end{center}
\label{tab:glob}
\end{table}

\section{Properties}\label{sec:properties}

\subsection{Age and Distance}\label{sec:age}

To determine the age of the object, we collect the sample of objects within 5$^{\circ}$ of ($l$, $b$) = (80.6, 26.1) that also fall into the dashed selection box in Figure \ref{fig:fehrv}. Using the spectroscopic data from the LAMOST pipeline, along with dereddened Two Micron All Sky Survey photometry (2MASS, \citealt{skr2006}; dereddened using the maps of \citealt{sch1998} and the filter coefficients of \citealt{dav2014} which are calculated relative to the $r_{SDSS}$ band extinction coefficient of \citealt{sch2011}), we fit these data to a series of Padova isochrones (\citealt{bre2012}, \citealt{che2014}, \citealt{tan2014}) with the metallicity set to the average of the candidate objects ([Fe/H] = -0.64).
Fitting is accomplished by finding the age of the isochrone which minimizes the $\chi^{2}$ value:

\begin{equation}
\chi^{2} =\left( \frac{T_{i} - T_{j}}{ \sigma_{T_{i}}}\right)^{2} + \left( \frac{log(g)_{i} - log(g)_{j}}{ \sigma_{log(g)_{i}}}\right)^{2},
\end{equation}

For each observed point $i$ and each isochrone $j$, where $T$ is the effective temperature and $log(g)$ is the surface gravity.

Once an age is selected, the distance to the population is estimated by minimizing the total euclidean distance of the candidates to the isochrone in temperature, surface gravity and magnitude space for varying distance moduli; where all coordinates are normalized over the range of that parameter in the isochrone.

\begin{figure}
\includegraphics[width=\linewidth]{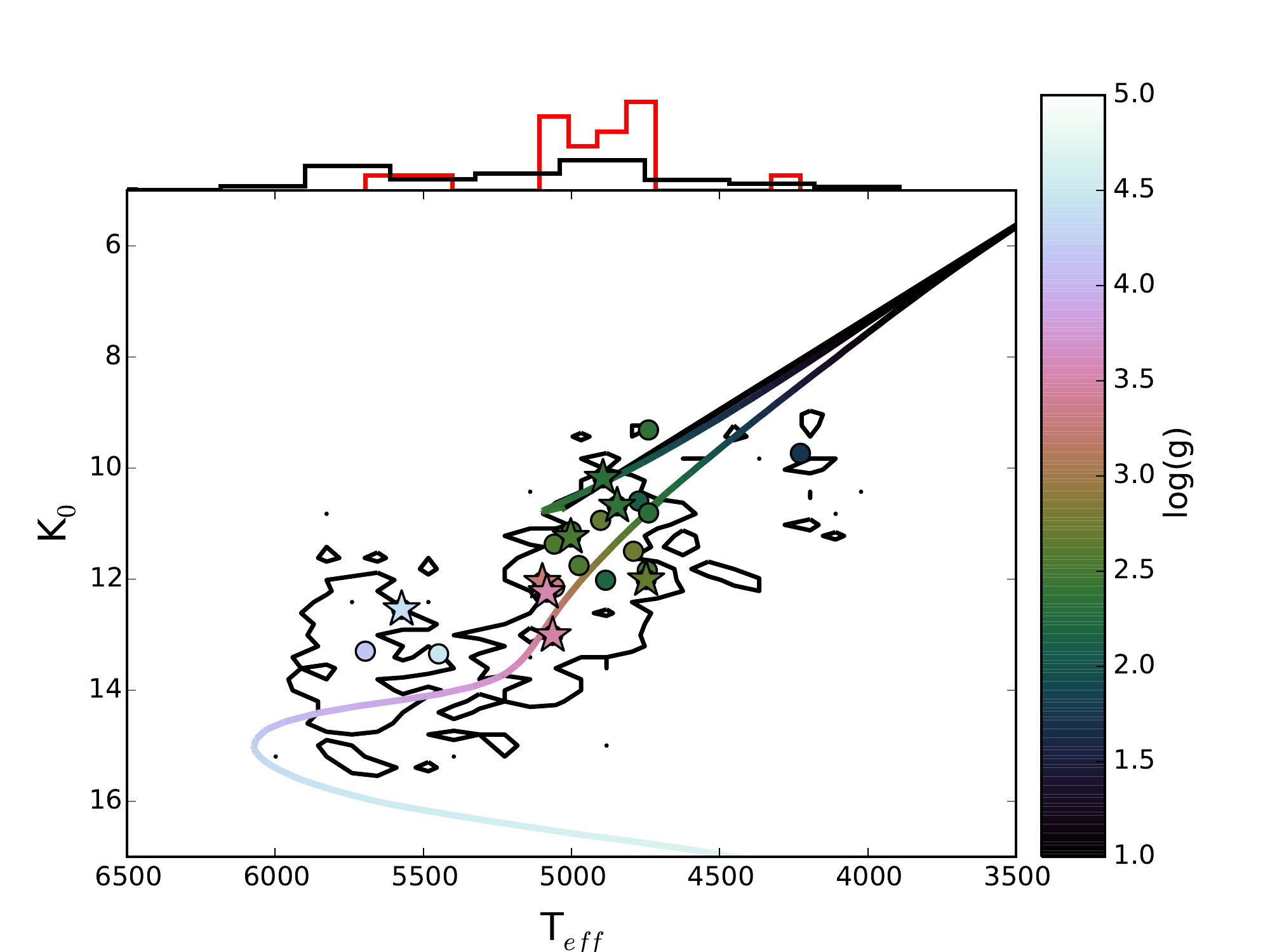}
\caption{Best fit Padova isochrone for the objects falling within 5$^{\circ}$ of ($l$, $b$) = (80.6, 26.1) and in the dashed selection box (circles) or solid selection box (stars). The contours indicate the locations of all stars in the LAMOST survey falling into the dashed selection box. This isochrone is 11 Gyr at a distance of 2.6 kpc. The normalized histograms above show the number of objects as a function of temperature for the target objects in red and the contoured objects in black. An overdensity of giants in the target selected stars relative to the all-sky sample can be noticed.}
\label{fig:iso}
\end{figure}

We find that the best fitting isochrone is 11 Gyr old (90\% probability of being $>$4.4 Gyr old) at a distance of 2.6 kpc. This fit is shown in Figure \ref{fig:iso}. We note that the distances to the stars independently estimated by the spectra (\citealt{wan2015}; \citealt{car2015}) are in good agreement (see Figure \ref{fig:schlaufman}). The normalized histogram above Figure \ref{fig:iso} shows the distribution of all stars which fall into our dashed selection box. The overdensity of cool giants in this part of the sky, relative to the general distribution of stars in this selection box, is further evidence that our detection is real.

\subsection{Velocity and Orbit}\label{sec:orbit}

The clump is quite tight in radial velocity dispersion, 9.0 km s$^{-1}$; however, our selection of candidate stars is constrained by a hard cut in velocity range, so this may not be indicative of the actual velocity dispersion. The distribution of UCAC4 proper motions are plotted in Figure \ref{fig:schlaufman}. 

Taking the average properties of the constituent stars, we calculate the orbit of the clump through the axisymmetric potential of \citet{deh1998} (potential 2b, chosen by comparison with \citealt{mcm2011}). We appear to be observing the clump close to its apogalacticon position at R=8.0 kpc. The orbit is fairly eccentric with a perigalacticon distance of 0.78 kpc and has a low inclination, with a maximum distance from the plane of 1.3 kpc.

The proper motion dispersion of the candidate stars at the distance of the cluster implies a much greater tangential velocity spread than radial velocity spread. If, as the orbit suggests, we are viewing the apocenter of the stream at a tangent to the orbit, we may expect to find a large proper motion dispersion. However, we suspect that this large dispersion is largely an effect of proper motion measurement uncertainty and field contamination.

A large fraction of these stars are observable at magnitudes of V $<$ 15, an observational space where the Gaia satellite will provide 12 $\mu$as yr$^{-1}$ proper motion errors. This is a factor of a hundred smaller than the proper motion errors available presently and would greatly improve our knowledge of the kinematic profile of this object.

\subsection{Observational Size, Mass and Luminosity}

From the independent distance estimates of \citet{wan2015}, the clump has a standard deviation in distance of $\sim$1.2 kpc; when considering the reported distance errors of $\sim$ 0.9 kpc, this is an actual depth of $\sim$910 pc. For a rough estimate of the angular extent, we look at the frequency of stars falling into our metallicity-velocity window as a function of distance from the center (see Figure \ref{fig:fehrv}). We find that the frequency falls to near-background levels at $\sim$4$^{\circ}$ in radius, which, at the distance estimated by the isochrone fit of 2.63 kpc corresponds to a width of 367 pc. Estimating the width with certainty is difficult without uniform photometric data, owing to the patchiness of the LAMOST coverage. However, the width and depth are similar to each other and the sightly more radial extent is consistent with the orientation of the orbit (i.e. viewing the apocenter at a tangent).

We did attempt to investigate the surface density of M-Giant stars in Wide-field Infrared Survey Explorer \citep{wri2010} data (see \citealt{li2015}), and 2MASS data as a way to characterize the spatial dimensions, but our overdensity was not found in those data.

We next make a crude estimate of the mass and luminosity of this object by integrating the functions:

\begin{equation}
mass = K \int m(J) \cdot IMF_{num.}(J) \cdot dJ,
\end{equation}
\begin{equation}
luminosity = K \int L(J) \cdot IMF_{num.}(J) \cdot dJ,
\end{equation}

over the total 2MASS J band ranges of the isochrones. L(J) and m(J) are estimates of the luminosity and mass of a star given a J band magnitude, interpolated from the isochrone. The IMF is the \citet{cha2001} initial mass function. K is determined to satisfy the equation:

\begin{equation}
N = K \int^{14}_{10} c(J) \cdot IMF_{num.}(J) \cdot dJ,
\end{equation}

where N=22$\pm\sqrt{22}$ is the number of stars we observe in the population, $c(J)$ is a the completeness of the LAMOST spectroscopy in the relevant color range, as estimated by 2MASS photometry.  This rough estimate places the total mass at $2.1\pm0.4\cdot10^{4}$ M$_{\odot}$ and the luminosity around $1.2\pm0.3\cdot10^{4}$ L$_{\odot}$.

\subsection{Progenitor}

The system's orbit, passing so deep into the Galactic potential, implies that an object of mass $2.1\pm0.4\cdot10^{4}$ M$_{\odot}$ would be tidally disrupted on relatively short timescales compared to the estimated age of 11 Gyr. However, it is known that structures remain coherent in velocity space long after being physically dissociated in the Milky Way potential \citep{hel1999}. The stream is also observed close to its calculated  apogalacticon, where stream members are known to pile up in physical space \citep{deh2004}.

\begin{figure}
\includegraphics[width=\linewidth]{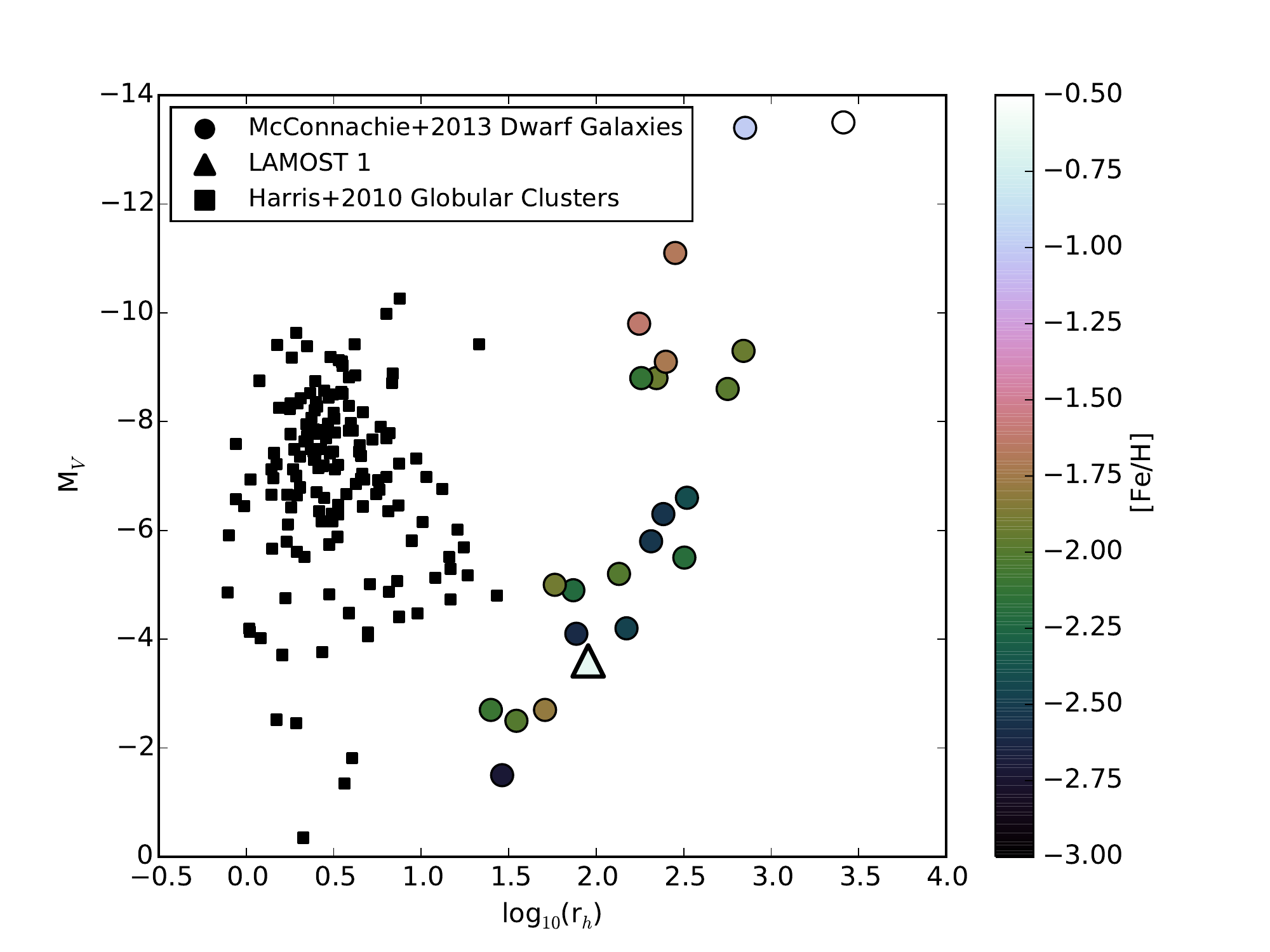}
\caption{Integrated V band magnitude as a function of half light radius for globular clusters from the catalog of \citet{har1996}, 2010 edition, and dwarf galaxies from the catalog of \citet{mcc2012}. The location of the clump discovered in this work is plotted and is seen to lie in the region inhabited by dwarf galaxies; however, it's metallicity (indicated by color) is seemingly incongruous with similarly sized dwarf galaxies. The clump we observe is likely just a portion of a larger, tidally disrupted satellite. Since we do not see strong spatially coherent structures, we do not think the progenitor likely to be a dwarf galaxy; since the orbit is relatively eccentric, we do not expect it to have originated from an open cluster.}
\label{fig:lumirad}
\end{figure}

The stream's progenitor could be either a dwarf galaxy or a star cluster. In Figure \ref{fig:lumirad} we compare the half light radius and bolometric magnitude of the clump, a collection of globular clusters taken from \citet{har1996}, and a collection of dwarf galaxies taken from \citet{mcc2012}. The high metallicity of our stream implies that, if the progenitor were a dwarf galaxy, it should be relatively massive, on par with that of the Sagittarius Dwarf. Since we do not see any prominent physical streams in the photometric data, we look to another possible explanation. If instead the progenitor is a star cluster, the elliptical orbit and advanced age of our stream argues toward a globular cluster origin.

\section{Discussion}\label{sec:discussion}

We have performed a gridwise search of LAMOST data for clumps in [Fe/H] and radial velocity space. Most of the algorithmic detections belong to known structures such as the Sagittarius tidal streams and the galaxy M13. A number appear to be spurious detections. However a significant detection has been noticed moving away from the sun at $\sim$164 km s$^{-1}$ at a metallicity of -0.64 dex.

We collect the believed constituents of this clump and compare to the metallicity-velocity, proper motion, and distance distribution of stars in a local background field, we find that this sample is a 5.1$\sigma$ signal over the background\footnote{We have also compared these objects to various background fields, including a Galactic mirror field, and with tighter and looser selection criterion, and the overdensity remains significant.}.

Fitting an isochrone to the stars, yields a best fit isochrone at 2.6 kpc that is most likely around 11 Gyr old and unlikely to be younger than 4.4 Gyr. Considering our observational completeness along with this best fit isochrone, we estimate the clump to have a mass of 2.1$\pm0.4\cdot10^{4}$ M$_{\odot}$ and a bolometric luminosity of 1.2$\pm0.3\cdot10^{4}$ L$_{\odot}$. The properties are summarized in Table 1. Note that since this clump is likely merely a portion of a larger stream, these are just estimates of the progenitors mass and luminosity.

The large physical size and the fact that the tangential velocity dispersion is so much higher than the radial velocity dispersion imply that this could be part of a tidal stream, although some tangential velocity dispersion is certainly an effect of proper motion uncertainty at this distance. Integrating its position and velocity through a galactic potential implies that we are observing the clump near its apogalacticon on an eccentric orbit. The high metallicity could point to a large dwarf galaxy progenitor, however, the lack of a strong physical overdensity argues against this interpretation. The above properties lead to our classification of this comoving clump as a disrupted globular cluster, which we dub Lamost 1.

\section{Acknowledgements}

We thank Dr. Emma Small for useful discussion.

We thank the developers and maintainers of the following software libraries which were used in this work: Topcat \citep{tay2005}, Aladin \citep{bon2000}, NumPy, SciPy, AstroPy \citep{ast2013}, astroML \citep{van2014}, matplotlib and Python.

Guoshoujing Telescope (the Large Sky Area Multi-Object Fiber Spectroscopic Telescope LAMOST) is a National Major Scientific Project built by the Chinese Academy of Sciences. Funding for the project has been provided by the National Development and Reform Commission. LAMOST is operated and managed by the National Astronomical Observatories, Chinese Academy of Sciences.

JJV acknowledges the support of a LAMOST fellowship. M.C.S. acknowledges financial support from the CAS One Hundred Talent Fund and from NSFC grants 11173002 and 11333003. This work was also supported by the Strategic Priority Research Program The Emergence of Cosmological Structures of the Chinese Academy of Sciences, Grant No. XDB09000000 and the National Key Basic Research Program of China 2014CB845700.

\end{document}